
%
%
%
%

 \input phyzzx
 \PHYSREV
 \doublespace
 \sequentialequations

\Picture\figA \height=2in \width=\hsize
\caption{\singlespace
Perturbative QCD mechanisms for generating
rapidity gap events. The dashed lines indicate
that the produced partons are in color-singlet state.
(a) Two final-state quark-antiquark pairs.
(b) A quark-antiquark jet-pair
and a two-gluon jet-pair.
}
\savepicture\figApic

\Picture\figB \height=2in \width=\hsize
\caption{\singlespace
A QED mechanism for generating
rapidity gap events. The variables $x_1$
and $x_2$ are the momentum fractions
carried by the quark inside each
jet-pair. The dashed lines indicate that
the produced quark and antiquark are
in color-singlet state.
}
\savepicture\figBpic

\Picture\figC \height=2in \width=\hsize
\caption{\singlespace
Rapidity gap event rate per million $Z$
in the symmetric gap cut case, for QCD and
QED.
}
\savepicture\figCpic

\Picture\figD \height=2in \width=\hsize
\caption{\singlespace
Rapidity gap event rate per million $Z$
in the asymmetric gap cut case, for two different
values of jet-pair invariant-mass cut ($M_c=30$ GeV
and $M_c=15$ GeV). The dashed lines are the event rates
for QCD-induced gap events, and the solid lines are
for QED-induced gap events.
}
\savepicture\figDpic

\def\bar#1{\overline{#1}}
\def\doeack{\foot{
  Supported in part by Department of Energy contract DE--AC03--76SF00515
  (SLAC) and contract DE--FG02--93ER--40762 (Maryland), and
  by the United Kingdom Science and Engineering Research Council
  (Durham).}}
 \Pubnum{SLAC--PUB--6113 \cr
DOE/ER/40762-009 \cr
U of Md. PP \#93-198}
 \date{May 1993}
 \pubtype{T/E}
 \titlepage
 \title{QED-Induced Rapidity-Gap Events at the Z Peak
 \doeack}
 \author{Hung Jung Lu}
 \address{Department of Physics, University of Maryland \break
  College Park, Maryland 20742, U.S.A.}
 \author{Stanley J. Brodsky}
 \address{Stanford Linear Accelerator Center \break
  Stanford University, Stanford, California 94309, U.S.A.}
 \andauthor{Valery A. Khoze}
 \address{Department of Physics, University of Durham \break
  Durham DH1 3LE, ENGLAND}
\vfill
\eject

 \abstract
We study rapidity-gap events in $e^+e^-$ annihilation at the Z boson peak
initiated by the emission of a virtual photon.
This mechanism is suppressed by the QED coupling constant,
but it is enhanced due to a large propagator term from the virtual photon.
For typical kinematics,
we find a smaller event rate than analogous QCD type gap events.
In the small jet-pair invariant mass limit,
the QED type events
follow a $1+\cos^2\theta$
distribution in the jet-pair scattering angle,
instead of the $\sin^2\theta$ distribution of the QCD case.

\vfill
 \submit{Physics Letters B}
\vfill
 \endpage

\REF\BBL{J. D. Bjorken, S. J. Brodsky and H. J. Lu,
            Phys. Lett. {\bf B286}, 153 (1992).}

\REF\Randa{J. Randa, Phys. Rev. {\bf D21}, 1795 (1980).}

\REF\MC{Monte Carlo simulations of QCD rapidity gap events
and their backgrounds at the SLC are now underway.
(P.N. Burrows, private communication.)}

\REF\Khoze{
        Yu.L. Dokshitzer, V.A. Khoze and S.I. Troyan, in
        Proc. Sixth Intern. Conf. on Physics in Collision
        (1986), ed. M. Derrick (World Scientific, Singapore, 1987)
        p. 417.
        Yu.L. Dokshitzer, V.A. Khoze and T. Sj\"ostrand,
              Phys. Lett. {\bf B274}, 116 (1992). }

\REF\BJ{J. D. Bjorken,  Phys. Rev. {\bf D47}, 101 (1992),
        Int. J. Mod. Phys. {\bf A7}, 4189 (1992). }

\REF\RG{
        H. Chehime, M.B. Gay Ducati, A. Duff, F. Halzen, A.A. Natale,
        T. Stelzer and D. Zeppenfeld,
              Phys. Lett. {\bf B286}, 397 (1992).
        R.S. Fletcher and T. Stelzer, BA--92--43 (1992).
        V. Del Duca and W.-K. Tang, SLAC--PUB--6095 (1993).
        H. Chehime and D. Zeppenfeld, MAD--PH--725 (1992),
                                      MAD--PH--748 (1993).
        E. Gotsman, E.M. Levin and U. Maor, TAUP--2030--93 (1993).
        E. Levin, FERMILAB--PUB--93--012--T (1993).}

\REF\ISR{V. N. Baier, V. S. Fadin, V. A. Khoze and E.A. Kuraev,
         Phys. Rept. {\bf 78}, 294 (1981).
         R. N. Cahn, Phys. Rev. {\bf D36}, 2666 (1987).
         J. P. Alexander, G. Bonvicini, P. S. Drell and
         R. Frey, Phys. Rev. {\bf D37}, 56 (1988).}

Jet events observed at $e^+e^-$ annihilation can be understood as
the creation of short-distance quarks and gluons which subsequently
materialize into hadrons. As the short-distance colored particles
move apart, the rapidity region separating the colored objects
is filled by the hadrons.
However, as pointed out in Refs. [\BBL,\Randa],
a perturbative QCD mechanism exists which
generates jet events containing rapidity gaps. The representative
diagrams are shown in Fig. \figA.
Essentially, two color-singlet collinear
jet pairs are produced at short distance. The production of hadrons
is expected to be suppressed in the rapidity gap region separating
the two color-singlet systems. This QCD mechanism is found to
give an observable rate at $Z$-peak for producing gap
events\refmark\MC.
Qualitatively, the rapidity gap events constitute a fraction
$$
 R^{\rm QCD}_{\rm gap} = \sigma^{\rm QCD}_{\rm gap}/\sigma_{\rm tot}
 \sim \alpha_s^2 {M_1^2\over s}\ {M_2^2\over s}
\eqn\eqA $$
of the total $Z$ cross section, where
$M_1^2$ and $M_2^2$ are the invariant mass of the jet pairs.

\midinsert\figApic\endinsert
\midinsert\figBpic\endinsert

The recent interest in the study of events
containing rapidity gaps has been motivated
by the use of these events as possible
triggering signals in high-mass scale physics [\BBL,\Khoze,\BJ,\RG].
In this article,  we shall study an interesting QED
mechanism that contributes additionally
to the generation of rapidity-gap
events in $e^+e^-$ annihilation at $Z$ peak. One
possibility is to replace the gluon propagator in Fig. \figA(a)
by a photon propagator. This contribution has an identical
kinematic dependence of the analogous QCD process, but it is
suppressed due to the smallness of the QED coupling constant.
In Fig. \figB\
we show the Feynman diagram in an alternative configuration,
where one of the color-singlet jet-pairs is effectively the
decay product of a virtual photon. There is no QCD analogy
for this configuration since the gluon is a color octet object.
Although the new contribution is still suppressed by $\alpha_{em}^2$,
the virtual photon propagator
can substantially compensate for the coupling constant suppression.
In fact, for small invariant-mass jet-pairs, the diagram in Fig. \figB\
acquires a large enhancement due to the small virtuality carried by
the photon. This should
be contrasted with the QCD diagram in Fig. \figA(a), where the quark
and antiquark produced by the virtual gluon are required to go
in opposite directions across the rapidity gap, so that
the virtuality carried
by the gluon is generally large.

 From power counting we expect the
QED rapidity gap events to be produced at a rate given by
$$
 R^{\rm QED}_{\rm gap} = \sigma^{\rm QED}_{\rm gap}/\sigma_{\rm tot}
 \sim \alpha_{em}^2 {M_2^2\over s}.
\eqn\eqB$$
That is, compared to the QCD case of Eq. \eqA, the event rate is enhanced by
a power of $s/M_1^2$, despite the suppression in the coupling
constant. In fact, for typical kinematics, the QED contribution
can become as large as the $q\bar qgg$ contribution
of the QCD case (Fig. 1(b)). (In Ref. [\BBL]
we have shown that the $q\bar qq\bar q$ contribution
of Fig. 1(a) is about an order of magnitude larger than
the $q\bar qgg$ contribution.)
In principle,  the QED diagrams interfere with the
QCD diagrams when the quark jets have identical flavor.
However, as we shall see, in the small invariant mass limit,
the interference effects become unobservable when we integrate
over the azimuthal angle of the individual jets with respect to
the thrust axis of the jet-pairs. In particular, this interference
has no effect on the total cross section. The QED and QCD
contributions to the event rate can therefore
be calculated separately.

As in Ref. [\BBL], we will neglect quark masses and consider
the small jet-pair invariant mass limit: $M_1^2, M_2^2 \ll s$. The
kinematic variables are as specified in Fig. \figB,
where $P_1$ and $P_2$
are the four momentum of the jet pairs,
and $x_1$, $1-x_1$, $x_2$, $1-x_2$ are
the momentum fractions of the individual jets inside the jet pairs.
We define the two-component weak charge of a fermion to be\refmark\BBL
$$
 {\bf Q}_f =
 \left({Q_f^L \atop  Q_f^R}\right)
 =
 \left({\sec\theta_{\rm w} I_f
      - \sin\theta_{\rm w} \tan\theta_{\rm w} Q_f \atop
      - \sin\theta_{\rm w} \tan\theta_{\rm w} Q_f}
 \right),
\eqn\eqC$$
where $\theta_{\rm w}$ is the weak angle, $I_f$ the isospin and $Q_f$ the
electric charge of the fermion $f$.
Using this notation, the total $e^+e^-$ annihilation
cross section around the $Z$ resonance can be
conveniently expressed as:
$$
\sigma_Z =
{\pi \over 3}
{ {\bf Q}_e^2 {\bf Q}_Z^2 \alpha^2_{\rm w} ~ s
  \over
  (s-M_Z)^2 + \Gamma_Z^2 M_Z^2
},
\eqn\eqCA$$
where
$$\eqalign{
 {\bf Q}_Z^2
&= \sum_f {\bf Q}_f^2 =
 \sum_l {\bf Q}_l^2 + 3 \sum_q {\bf Q}_q^2 \simeq 3.771
\cr
 ( l
&= e, \mu, \tau; q = u, d, c, s, b ),
\cr
 {\bf Q}_f^2
&= {Q_f^L}^2 + {Q_f^R}^2 ,
\cr
 \alpha_{\rm w}
&={ g_{\rm w}^2 \over 4 \pi} = {e^2 \over 4 \pi \sin^2\theta_{\rm w}}
 \simeq {1 \over 29.3},
\cr
 M_Z, \Gamma_Z
&= {\rm ~mass~and~width~of~the}~Z~{\rm boson.}
\cr
} \eqn\eqCB $$
We shall later use this cross section to normalize the
production rate of rapidity-gap events. This is subject
to the same caveat pointed out in Ref. [\BBL]. Namely,
initial state radiation induces a substantial correction to the above
result for $\sigma_Z$ [\ISR]. However, the same effect
is present in rapidity-gap events; thus we expect
these effects to largely cancel when we consider
ratios of cross sections.

There are three additional
Feynman diagrams contributing to QED
rapidity gap events similar to the one depicted in Fig. \figB.
Basically, there are two diagrams with the virtual photon
decaying into $q_a {\bar q}_a$ and another two diagrams
with the virtual photon decaying into $q_b {\bar q}_b$.
The intermediate formulas in the calculation of the helicity
amplitudes are complicated and will not
be presented here. However, we shall display
a particular helicity amplitude in order to facilitate
later discussion. Let us momentarily add up the contributions
of the  two Feynman
diagrams with the virtual photon decaying into $q_a {\bar q}_a$
and designate this sum by $i {\cal M}^a$.
Keeping only leading contribution in the jet-invariant mass $M_1$, we
obtain the following amplitude
for all-positive fermion helicities (one right-handed
electron line and two right-handed quark lines):
$$\eqalign{
i {\cal M}^a
(+++) & =
{4 g_{\rm w}^2 e^2 Q^R_e Q^R_a Q_a Q_b \sqrt{s}\over
     (s-M_Z^2+i \Gamma_Z M_Z) M_1} \cr\crr
&\quad
\left\{ (1-x_1)
        \sqrt{{x_2\over 1-x_2} }
        \cos^2(\theta/2) e^{i \phi_1}
      - x_1
        \sqrt{{1-x_2\over x_2} }
        \sin^2(\theta/2) e^{-i \phi_1}
\right\}, \cr}
\eqn\eqD  $$
where $g_{\rm w}$ is the weak coupling
constant, $\theta$ is the
polar angle of the jet-pair thrust axis, and $\phi_1$ is the azimuthal
angle of the $q_a$ jet with respect to the thrust axis. The sum of
the other two amplitudes where the virtual photon decays into
$q_b \bar{q}_b$ can be obtained by interchanging
$a\leftrightarrow~b$, $x_1\leftrightarrow~x_2$,
$\phi_1\leftrightarrow~\phi_2$ in the previous formula. (We will
refers to this sum as $i {\cal M}^b$.) Other helicity
amplitudes can be deduced from the previous formula by various
conjugation operations, but for simplicity we shall omit them here.
Notice the presence of the $e^{\pm i \phi_1}$ dependence. This
dependence on the azimuthal angle $\phi_1$ is absent for the QCD
amplitudes as those in Fig. \figA. Therefore, the interference
effect between the QED and QCD amplitudes disappear upon
integration of the $\phi_1$ angle. Similarly, the interference
term between $i {\cal M}^a$ and $i {\cal M}^b$ disappear after
integrating out $\phi_1$ or $\phi_2$. In short, interference
effects only alter the azimuthal angle distribution of the
individual jets with respect to the thrust axis, and these effects
become unobservable when we integrate out those azimuthal angles.

The difference in the azimuthal angle dependence
in QED and QCD is related the following fact.
In the QCD case, the outgoing parton pairs
can be perfectly aligned. In this
collinear limit the azimuthal angles of the individual
jets with respect to the thrust axis
are not expected to play a role in the scattering amplitude.
In the QED case, the jets originated from the virtual photon are
forbidden to be exactly parallel, since a
timelike virtual photon with helicity $\pm 1$ cannot
decay into two parallel, massless quarks:
some non-collinearity is required in order to
carry the photon's polarization.
The appearance
of the azimuthal dependence reflects the correlation
between the outgoing jets with the event plane
formed by the virtual photon and the beam direction.

Considering only $i {\cal M}^a$, after squaring, averaging and adding the
various helicity contributions, counting the color multiplicity
of the quarks, symmetrizing the momentum fraction variables,
integrating out the azimuthal angle $\phi_1$,
and normalizing the cross section with respect
to the total $Z$ cross section, we obtain
$$\eqalign{
{\sigma_a\over \sigma_Z}
&= {27\over 32}\
    \left({\alpha_{em}\over \pi}\right)^2
    {Q_a^2 Q_b^2 {\bf Q}_b^2 \over
         {\bf Q}_Z^2}\
\int
    {d M_1^2\over M_1^2}\
    {d M_2^2 \over s}\
    d x_1\, d x_2\, d \cos\theta
\cr\crr
&\quad\times
    \left[ x_1^2 + (1-x_1)^2
    \right]
    \left[ {x_2\over 1-x_2} +
           {1-x_2\over x_2}
    \right]
    (1+\cos^2\theta)\ . \cr}
\eqn\eqE $$
Notice first the $1+\cos^2\theta$ term in the above formula.
This angular dependence differs from the $\sin^2\theta$
distribution obtained for QCD induced gap events. In particular,
the QED type events can become the dominant contribution
for polar angles near the forward and backward beam direction.
Notice also the integral of $M_1^2$ has an apparently infrared
divergence when $M_1^2 \to 0$. In reality this divergence does
not occur due to the physical energy threshold $\widetilde M_a$
for the production of the quark-antiquark pair $q_a \bar q_a$
from a virtual photon. Numerically we will take
$\widetilde M_a = M_\rho,\ M_\omega,\ M_\phi,\ M_{J/\psi}$ and
$M_\Upsilon$ as representative values for the threshold energy
of producing quark-antiquark pairs of flavor
$u,\ d,\ s,\ c$ and $b$.
Finally, notice
the factorization of the $x_1$, $x_2$ and $\cos\theta$
dependence. Eq. ~\eqE ~ can therefore be interpreted as the probability
of the radiative decay of $Z$ into a quasi-collinear
quark-antiquark pair and a virtual photon, multiplied
by the probability of a photon ``splitting" into a
quasi-collinear quark-antiquark pair (represented by
the $x_1^2+(1-x_1)^2$ term in the equation). The singularity
in the $x_2$ and $1-x_2$ denominators reflects the
infrared divergence in the virtual quark propagator
when its associated quark or antiquark becomes soft
(See Fig. 2).
However, as in the QCD case [\BBL,\Randa], when we impose the existence of
a rapidity gap, the virtual photon and the other jet-pair
are required to go  in opposite direction across the rapidity
gap, hence the momentum transfer of the virtual quark
is generally hard and the infrared divergence is avoided.

Taking now into account contributions from $i {\cal M}^b$,
adding over all quark flavor combinations,  symmetrizing
the identical flavors cases, and integrating over $\theta$,
we obtain
$$\eqalign{
R^{\rm QED}_{\rm gap}
&=
\left( {9\over 4}
\right)^2
\left({\alpha_{em}\over \pi}
\right)^2
{\sum Q_b^2 {\bf Q}_b^2\over {\bf Q}_Z^2}
\sum Q_a^2
\int_{\widetilde M_a^2}
{d M_1^2\over M_1^2}
\int
{d M_2^2\over s}  \cr\crr
&\quad \times
\int d x_1 \left[ x_1^2 + (1-x_1)^2
           \right]
\int d x_2 \left[ {x_2\over 1-x_2} +
                  {1-x_2\over x_2}
           \right]\ . \cr}
\eqn\eqE $$
In principle, we should consider also the
QED contribution from diagrams like Fig. \figA(a) where
the gluon propagator has been replaced by a photon propagator.
These diagrams contribute mainly through interference
effects with the corresponding QCD diagrams.
A detailed analysis reveals that these effects are negligible.
(The resulting interference
causes only a 0.6\% decrease in the QCD-type event rate for the
symmetric gap case.)

The limits of the various integrals in the above formula
depend on the physical
cuts we impose for the selection of events.
As in Ref. [\BBL], we first analyze the event rate for
a symmetric gap cut case and then repeat the analysis for
the asymmetric gap cut case. In the former case, we sum over
all events with all the produced jets having an absolute
rapidity greater than $g/2$ with respect to the jet-pair
thrust axis.
(This is subject to the same caveat pointed out in Ref. [\BBL];
namely, due to the effect of hadronization process, the
hadron fragments of each quark are concentrated within a circle
of radius $\sim 0.7$ in the lego plot. The physically observed
gap is thus expected to have a width $g_{\rm eff} \sim g - 1.4$.)

\midinsert\figCpic\endinsert
\midinsert\figDpic\endinsert

The QED and QCD type gap event rate per million Z events,
is shown in Fig. \figC\
as a function of the symmetric rapidity gap cut $g.$
In Fig. \figD\ we plot the event rate per million Z events
when the gap is not required to be symmetric and the
jet-pair invariant masses are required to be less than
$30$ or $15$ GeV.
We have used a value $\alpha_s = 0.13$ for the
strong coupling constant.
We see from these figures that in the large
rapidity gap region the QED-induced events can constitute
a substantial fraction of the QCD type events.
For instance, $R^{\rm QED}_{\rm gap}/R^{\rm QCD}_{\rm gap}
\sim 0.11$ for $g=4$ in the symmetric gap case.
The event rate for a larger gap region is probably too small for
present experimental observation.
In Ref. [\BBL] we have pointed out that in the QCD case,
the $q\bar qgg$ type events are suppressed by a factor $0.159$
with respect to the $q\bar qq\bar q$ type events in the
symmetric gap case.
(This relative suppression can be understood on the basis
of color factors:  at large $N_C$ the rate for color singlet
production $(q \bar q) + (gg) $ is proportional to $N_C,$
whereas the rate for two mesonic dijets $(q \bar q)$ $(q \bar q)$
is of order $N^2_C.$)
Hence the QED type events can become
as important as the $q\bar qgg$ type rapidity gap events in the large
gap region ($g\ge 4$).

In summary, we have studied QED mechanisms for
producing jet event containing large rapidity gaps
in $e^+e^-$ annihilation at the $Z$ peak energy. For typical
kinematics, the event rate is found to be small compared
to the rate for the corresponding QCD processes. However,
the QED-induced  rapidity gap events have some distinctive features.
For instance,
the QED events are distributed as $1+\cos^2\theta$ in terms
of the polar angle $\theta$ of the thrust axis
as opposed to the QCD events which are distributed
as $\sin^2\theta$.
The QCD and QED mechanism also favor different
flavor combinations. A process involving
a $(b\bar b)$  and  $(c\bar c)$ color singlet jet-pair
is clearly absent in the QCD $q\bar qq\bar q$ mechanism.
Also, in the QED case, one of the jet-pairs tends to
have a small invariant mass. In fact,
the virtual photon easily transforms
into its hadronic components ($\rho, \omega, \phi, J/\psi, \Upsilon$).
Thus, a likely signal of the QED mechanism would be a
vector meson going in one direction and a hadronic
system going in the opposite direction, with a large rapidity
gap in between. Events
containing a narrow resonance such as the $J/\psi$ recoiling against a
dijet system could be particularly interesting since the direction and
polarization of the vector meson is revealed through its decay into a
lepton pair.
Combining with beam polarization, the detailed study
of these events can offer non-trivial
tests of Standard Model features.
Monte Carlo study of the hadronization
stage [\MC] would allow us to distinguish the gap events
produced through the described mechanisms from
the random fluctuation of hadron fragments.
Finally, let us point out that
at future linear colliders (energies above the $Z$ mass)
the QED-induced rapidity gap events can also
come from configurations corresponding to the
``radiative tail" of the $Z$. That is, a low
mass photon emitted in the initial state
takes away the right amount of energy to
``restore" the virtual $Z$ into resonance, and
two color singlet jet-systems are generated separately
as the decay products of the virtual photon
and the resonant $Z$.

\ACK

We thank James D. Bjorken for helpful discussions.

\refout
\end